\documentclass[12pt,a4paper]{article}
\newcommand{\eqref}[1]{(\ref{#1})}
\usepackage{graphics}
\usepackage{epsfig}
\begin{document}
\baselineskip18pt
\begin{center}
{\Large {\bf QHJ, WKB and exact quantisation} }\\
S. Sree Ranjani$^{a}$, A. K. Kapoor$^b$ and P.K. Panigrahi$^c$\\
$^{a,b}$School of Physics, University of Hyderabad, Hyderabad 500 046, India.\\
$^c$Indian Institute of Science Education and Research (IISER) Kolkata, Salt
Lake, Kolkata 700 106, India.\\
$^c$Physical Research Laboratory, Navrangpura, Ahmedabad - 380 009, India
\end{center}
\begin {center}
{\bf Abstract}
\end{center}
\indent
 We present a simple derivation of the WKB quantisation condition using the quantum Hamilton-Jacobi formalism and propose an exact quantisation condition within this formalism for integrable models in higher dimensions.

\noindent
\section{Introduction}
 The quantum Hamilton-Jacobi (QHJ) formalism,
developed in its present form  by Leacock and Padgett \cite{lea1}, \cite{lea2},  formally maps on to the classical Hamilton-Jacobi theory \cite{gold} in the limit $\hbar \rightarrow 0$. Just as the classical theory allows one to obtain the    frequency of the periodic
motion, without explicitly finding the solution of the equation of
motion, its quantum counterpart, the QHJ
theory, allows
one to find the energy eigenvalues without solving  the
Schr\"odinger equation.  An important role is played by the exact
quantisation condition in one dimension,
\begin{equation}
\frac{1}{2\pi}\oint_C pdx = n\hbar,  \label{exq}
\end{equation}
where the contour $C$ in the complex $x$-plane encloses the classical region between the two turning points $x_1$ and $x_2$ as shown in Fig. 1. In the above  equation $p$ is the quantum momentum function (QMF) defined  by
\begin{figure}[ht]
\centering
\epsfig{file=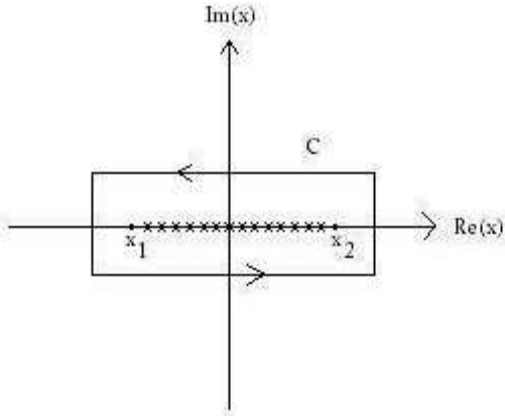,scale=1}
\caption{The contour $C$ enclosing the poles of the QMF $p$, corresponding to the nodes of the wave function $\psi$ in between the turning points $x_1$ and $x_2$.}
\end{figure}
\begin{equation}
p=-i\hbar\frac{d}{dx}\ln \psi,   \label{qmf}
\end{equation}
where $\psi$ is the eigenfunction of the Hamiltonian.

It should be emphasized that the above quantisation
   condition is well formulated for problems in one  dimension and for
   separable problems in higher dimensions. This rule has been successfully applied to different types of problems in one dimension which includes quasi-exactly solvable and periodic potentials. The procedure to get the eigenvalues, eigenfunctions and quantisation conditions on certain parameters in case of the quasi-exactly solvable problems are now well established. We briefly review these results in sec. 2.

   
  In the light of the QHJ equation,
\begin{equation}   
p^2 + i \hbar  p^{\prime}= 2m(E-V(x)),  \label{qhj}
\end{equation}
reducing to the classical HJ theory, $p= \sqrt{2m(E-V(x))}$,  in  $ \hbar \rightarrow 0$ limit, the Wentzel-Kramers-Brillouin (WKB) quantisation condition is expected to follow  from the exact quantisation condition as a natural consequence in the same limit. This derivation, shown in sec. 3, has a  simplicity and elegance which  can not be  matched by the text book treatment.

  The fact that the WKB quantization condition has a generalization to higher dimensions in the form of Einstein-Brillouin-Keller (EBK) quantization condition, motivates us to explore the precise origin of WKB quantization in the context of QHJ condition. This may reveal a route to appropriately extend  the exact quantization condition within the QHJ formalism to higher dimensions analogous to the EBK quantization condition. In sec. 4, we explain how Einstein's paths on the invariant tori used in the EBK quantisation condition and the QMF can be used to define  an exact quantization condition for integrable systems in  higher dimensions.  
   
   
\section{Applications of the Quantum Hamilton-Jacobi formalism}

 For a large number of potentials belonging to the classes of exactly solvable (ES) models, quasi-exactly solvable (QES) models \cite{ush}, periodic potentials and PT-symmetric potentials, exact expressions for energy eigenvalues and eigenfunctions follows from the QHJ formalism in a very simple and elegant fashion, making use of well known results in complex variables.

     The use of the quantisation condition \eqref{exq} for ES models gives exact expressions for the energy eigenvalues, whereas for the QES models, it gives the quasi-exact solvability  condition \cite{qes}. In comparison to the conventional method of obtaining solutions of a periodic potential \cite{arscot}, the QHJ method turned out to be rather simple, as demonstrated in the context of Lam\'e and the associated Lam\'e potentials. The former is ES, whereas the later is ES or QES depending on the potential parameters. For both the potentials, we obtained the band-edge eigenvalues and eigenfunctions along with the quasi-exact solvability  condition for the appropriate cases \cite{pp1}, \cite{sree3}. The fact that this method works in the complex plane made it a natural choice to analyse the PT-symmetric potentials, which are complex potentials with real eigenvalues. In addition, certain surprising results such as the exactness of SUSY WKB in the supersymmetric quantum mechanics become transparent when analysed using the QHJ formalism \cite{SUSYWKB}.

An investigation of certain potentials like the Scarf-I potential and the Scarf potential, which exhibit different behaviours for different potential parameters turned out be an interesting exercise. The broken and unbroken phases of SUSY in the case of Scarf-I and the band and  bound state spectrum in the case of the the Scarf potential \cite{scarf}, for different parameter ranges  are well distinguished. Suitable boundary conditions on the QMF, automatically separate the two parameter ranges and allow us to obtain the solutions in these two regimes in a straight forward fashion \cite{the}, \cite{anp}. 

\noindent
\section{WKB quantisation through QHJ formalism}
 In this section, we shall derive the WKB quantisation condition,
 \begin{equation}
   \oint p_{cl} dx= (n+\frac{1}{2})h.  \label{wkb1}
\end{equation}
 It is natural to formally explore the origin of WKB from the QHJ equation \eqref{qhj}.
 For this purpose, we first expand the QMF  in terms of $\hbar$ as
\begin{equation}
p=q_{0}+\hbar {q}_{1}+\hbar^2 {q}_{2}+\dots ,  \label{qn4}
\end{equation}
where $q_0,q_{1},$ etc., are functions of $x$. Substituting this in the QHJ
equation \eqref{qhj} and comparing the powers of $\hbar$ on both sides
gives,
\begin{eqnarray}
q_{0}&=&\sqrt{2m(E-V(x))},  \label{qo}\\
\qquad q_{1}&=&\frac{i}{2}\frac{q_{0}^{\prime}(x)}{q_{0}(x)}.   \label{q1}
\end{eqnarray}
Here \eqref{qo} is just the classical
momentum function $p_{cl}$. Thus, the QMF up to the first order in $\hbar$ is given by
\begin{eqnarray}
 p &\sim&p_{cl}+\frac{i \hbar}{2}\frac{d}{dx}\ln p_{cl}   \label{qmfh} \\
   &\sim& p_{cl} + \frac{i \hbar}{4} \frac{F\,^\prime}{F},
\end{eqnarray}
where $F(x)=2m(E-V(x))$, which vanishes at the turning points. The above is in complete analogy to the standard treatment of WKB approximation, where a detailed analysis leading to the connection formulae, involving Airy functions, becomes necessary to obtain the WKB quantisation rule.
We show that the same result follows when the above approximate expression for $p$ is substituted in the exact quantisation condition.
\begin{equation}
 \oint_C p_{cl} dx + \frac{i \hbar}{4} \oint_C \frac{F^{\prime}(x)}{F(x)}\,dx =n h.   \label{feqn}
\end{equation}
Assuming linear turning points and noting that the residue of   $F^\prime(x)/F(x)$ at each turning point is unity, an application of the
Cauchy residue theorem yields
\begin{equation}
\frac{i \hbar}{4} \oint_C \frac{F^{\prime}(x)}{F(x)}\, dx=-\frac{h}{2},
\end{equation}
which leads to the WKB quantisation condition
\begin{equation}
 \oint p_{cl} dx = (n+\frac{1}{2})h.  \label{wkb2}
\end{equation}
Here a transition, from the integral over a contour in the complex $x$-plane to an integral between the two turning points, has been made by a limiting process where the contour $C$ is shrunk to the real line interval between the turning points. It is important to note that the above derivation assumes the potential to be analytic on and inside the contour $C$, see Fig.1, except at the poles of the QMF corresponding to the nodes of the wave function $\psi$. This derivation of the WKB rule is not applicable to potentials with jump discontinuities.

 This illustrates the power and simplicity of the QHJ formalism.  This derivation of  WKB quantisation condition in the semiclassical limit prompts one to look for the possibility of obtaining an exact quantisation condition in higher dimensions.

\noindent
\section{A proposed exact quantisation condition for higher dimensions}
We now proceed to generalise \eqref{exq} and propose an exact quantisation condition
for integrable systems in higher dimensions. We begin by recalling
the Sommerfeld-Epstein quantisation condition,
\begin{equation}
 \oint p_{cl \,i} dq_i   = n_i h,  \label{sec}
 \end{equation}
where $p_{cl\,i}$ are the components of the classical momentum function and $i=1,2....d$, $d$ being the degrees of freedom. The line integral is along a closed periodic orbit as against the contour integral in the complex $x$-plane  in  \eqref{exq}.

Einstein pointed out that the quantisation condition \eqref{sec} is not canonically invariant and relies on the separability of the problem \cite{ein}. Noting that the separability has nothing to do with the actual quantum mechanical problem, Einstein  modified \eqref{sec} to
\begin{equation}
 \oint \sum _i p_{cl\,i} dq_i = n h, \label{inv}
 \end{equation}
which is canonically invariant, making the Sommerfeld-Epstein condition coordinate independent.  As observed by Einstein, the application of the above quantum rule \eqref{sec} requires the existence of paths, such that a single path determines a $p_{cl\, i}$ field for which a potential $J^{\ast}$ (Hamilton's characteristic function) exists:
\begin{equation}
p_{cl\,i}= \frac{\partial J^{\ast}}{\partial q_i}.   \label{p}
\end{equation}
These paths are located in the classically allowed region of the coordinate space and pass through each small neighbourhood an infinite number of times with only a finite number of different momentum directions. For such paths, the action $J^{\ast}$, will be a function on the $d$-torus and the gradient of the action, will be single valued.

Einstein further pointed out that the line integral in \eqref{inv} has the same value for all those paths which can be continuously transformed into each other. For all the curves, which by continuous transformation can be pulled together to a point, the integral \eqref{inv} vanishes. However if the space of $q_i$ is a multiply connected one, then there will be closed paths which cannot be  contracted to a point. Then if $J^{\ast}$ is an infinitely many-valued function, the integral in \eqref{inv} is non-zero for such paths.

The two important inputs for integrable models in Einstein's proposal have been the choice of the paths in the configuration space and the classical momentum being the gradient of a single potential function \cite{pankaj}, \cite{stone}\footnote{This paper gives a lucid elucidation of Einstein's insight into the problem of quantisation of non-separable systems. It also explains how Einstein unknowingly stumbled onto the concept of non-integrability and its relation with ergodicity, without actually realizing its implications for the area of dynamical systems.}. The requirement of  integrability ensures the existence of closed paths, as above, on the surface of the invariant torus in the phase space and the applicability of the Einstein quantisation for a given system. Also for integrable models, in general, there will be $d$- independent closed paths which cannot be deformed into each other and for each such path, \eqref{inv} gives an independent quantum condition.

We observe that, by virtue of its definition, QMF is always the gradient of the quantum action. As EBK quantisation condition naturally uses  Einstein's paths on the surface of the invariant tori, we suggest that the integration of the QMF along these paths can provide an exact quantization condition for integrable systems in higher dimensions. Therefore,  we propose  that the classical momentum function, $p_{cl\,i}$, in Einstein's condition, \eqref{inv}, be replaced with the QMF given by
\begin{equation}
p_i = -i\hbar\Big( \frac{\partial \log \psi}{\partial q_i}\Big),
\end{equation}
which leads to
\begin{equation}
 \oint \sum _i p_i dq_i = n h. \label{genqhj}
 \end{equation}
This gives us a  quantisation condition which is invariant under point transformations and which goes over to the condition proposed by Einstein as $\hbar \to 0$. We therefore propose this as a candidate for exact quantisation rule for integrable systems in higher dimensions. We would also like to point out that the above quantization condition gives correct results in the limit where the system becomes separable. In contrast to this the quantization condition proposed by Gutzwiller \cite{gut} fails to give correct results in the separable limit. Though Miller proposed a modified condition to rectify this problem, it gave correct results only in the limit that the system is a separable set of harmonic oscillators \cite{miller}.

To conclude, in this paper we have given a simple derivation of the WKB quantisation rule and have proposed an exact quantisation condition for integrable  systems in higher dimensions. Just as the WKB quantisation rule follows from the exact quantisation condition, we expect \eqref{genqhj} to go over to the EBK quantisation rule for integrable systems in the semiclassical limit\cite{ein, bri,kel}. We also point out here that the semiclasscal treatment of the billiard problem has been extensively investigated in the literature. The proposed quantisation condition can be tested for these models and the results can be compared with the existing results which will be reported elsewhere. 

\noindent
{\bf Acknowledgments} AKK and SSR thank A. P. Balachandran for helpful discussions.


\vspace{5mm}
\noindent
{\bf  References}

\begin{enumerate}

\bibitem{lea1} R. A. Leacock and M. J. Padgett, {\it Phys. Rev. D} {\bf 28}, 2491 (1983).

\bibitem{lea2} R. A. Leacock and M. J. Padgett, {\it Phys. Rev. Lett.} {\bf 50}, 3 (1983).

\bibitem{gold} H. Goldstein, {\it Classical Mechanics} (Addison-Wesley/Narosa, Indian student edition, 1995).

\bibitem{ush} A. Ushveridze, {\it Quasi-Exactly Solvable Models in Quantum Mechanics} (Bristol: Institute of Physics Publishing, 1994).

\bibitem{qes} K. G. Geojo, S. Sree Ranjani and A. K. Kapoor, {\it J. Phys. A: Math. Gen.} {\bf 36}, 4591 (2003).

\bibitem{arscot} F. M. Arscot, {\it Periodic Differential Equations} (Pergamon,Oxford, 1964).

\bibitem{pp1} S. Sree Ranjani, A. K. Kapoor and P. K. Panigrahi, {\it Mod. Phys. Lett. A} {\bf 19}, 2047 (2004).

\bibitem{sree3} S. Sree Ranjani, A. K. Kapoor and P. K. Panigrahi, {\it Int. jour. Theoretical Phys.} {\bf 44}, 1167 (2005).

\bibitem{pt} S. Sree Ranjani, A. K. Kapoor and P. K. Panigrahi, {\it Int. Jour. Mod. Phys. A.} Vol. 20, 4067 (2005).

\bibitem{SUSYWKB} R. S. Bhalla, A. K. Kapoor and P. K. Panigrahi, {\it Phys. Rev. A} {\bf 54}, 951 (1996).

\bibitem{susybook} F. Cooper, A. Khare and U. Sukhatme, {\it Supersymmetry in Quantum mechanics} (World Ecientific, Singapore, 2001).

\bibitem{scarf} F. L. Scarf, {\it Phys. Rev.} {\bf 112}, 1137 (1958).

\bibitem{the} S. Sree Ranjani, Thesis titled: {\it Quantum Hamilton - Jacobi solution
  for spectra of several one dimensional potentials with special
  properties} (Thesis submitted to the University of Hyderabad), Preprint quant - ph/0408036.

\bibitem{anp} S. Sree Ranjani, A. K. Kapoor and P. K. Panigrahi, {\it Ann. Phys.} {\bf 320}, 164 (2005).

\bibitem{ein} A. Einstein, {\it Verh. Deutsche. Phys. Gesell, Berlin}  {\bf 19}, 82 (1917).

\bibitem{pankaj} Pankaj Sharan, {\it Am. J. Phys.} {\bf 50}, 351 (1982).

\bibitem{stone} D. Stone, {\it Phys. Today}, {\bf 58}, 37 (2005).

\bibitem{gut} M. C. Gutzwiller, {\it J. Math. Phys.} {\bf 12}, 343 (1971).

\bibitem{miller} W. H. Miller, {\it J. Chem. Phys.} {\bf 63}, 996 (1975).

\bibitem{bri} L. Brillouin, {\it J. Phys. Radium}, {\bf 7}, 353 (1926).

\bibitem{kel} J. B. Keller, {\it Ann. Phys. (NY)}, {\bf 4}, 180 (1958).
\end{enumerate}
\end{document}